\documentclass[useAMS,usenatbib]{mn2e}

\usepackage{graphicx}
\usepackage{epstopdf}
\usepackage{url}

\newcommand\ion[2]{#1$\;${\scshape \lowercase{#2}}\relax}

\newcommand{\eg}{e.g.}

\newcommand{\fig}[1]{Fig.~\ref{#1}}

\newcommand{\kms}{km s$^{-1}$}
\newcommand{\logg}{$\log g$}

\newcommand{\msun}{$M_{\sun}$}

\newcommand{\teff}{$T_{\rm eff}$}
\newcommand{\vinf}{$v_{\infty}$}

\newcommand{\fuse}{{\em FUSE}}
\newcommand{\hst}{{\em HST}}

\begin{document}


\title[vZ 1128 in M3]{{\em FUSE}, STIS, and Keck spectroscopic analysis of the UV-bright star vZ 1128 in M3 (NGC 5272)}

\author[P. Chayer et al.]{P. Chayer,$^1$\thanks{E-mail: chayer@stsci.edu} 
W. V. Dixon,$^1$ 
A. W. Fullerton,$^1$
B. Ooghe-Tabanou,$^{2,3}$ and
I. N. Reid$^1$ \\
$^1$Space Telescope Science Institute, Baltimore, MD 21218, USA \\
$^2$\'Ecole Normale Sup\'erieure, Laboratoire de Radio-Astronomie, 28 rue Chomond, 75005 Paris, France\\
$^3$Current address: Sciences Po--M\'edialab, 84 rue de Grenelle, 75007 Paris, France\\
}

\maketitle

\label{firstpage}

\begin{abstract}
We present a spectral analysis of the UV-bright star vZ 1128 in M3 based on observations with the {\em Far Ultraviolet Spectroscopic Explorer (FUSE)}, the Space Telescope Imaging Spectrograph (STIS), and the Keck HIRES echelle spectrograph. By fitting the \ion{H}{I}, \ion{He}{I}, and \ion{He}{II} lines in the Keck spectrum with non-LTE H-He models, we obtain $T_{\rm{eff}} = 36$,600~K, $\log g = 3.95$, and $\log N({\rm{He}})/N({\rm{H}}) = -0.84$. The star's \fuse\/ and STIS spectra show photospheric absorption from C, N, O, Al, Si, P, S, Fe, and Ni. No stellar features from elements beyond the iron peak are observed.  Both components of the \ion{N}{V} $\lambda 1240$  doublet exhibit P~Cygni profiles, indicating a weak stellar wind, but no other wind features are seen.  The star's photospheric abundances appear to have changed little since it left the red giant branch (RGB).  Its C, N, O, Al, Si, Fe, and Ni abundances are consistent with published values for the red-giant stars in M3, and the relative abundances of C, N, and O follow the trends seen on the cluster RGB.  In particular, its low C abundance suggests that the star left the asymptotic giant branch before the onset of third dredge-up.  
\end{abstract}

\begin{keywords}
spectroscopy --- stars: abundances --- stars: individual: NGC 5272 vZ 1128
\end{keywords}

\section{Introduction}

In the color-magnitude diagrams of globular clusters, UV-bright stars are those objects bluer than the red giant branch and brighter than the horizontal branch.  They consist of stars that are evolving to the white dwarf stage, either from the asymptotic giant branch (AGB) or directly from the extreme horizontal branch (EHB).  Their atmospheric parameters and abundances should thus provide important constraints on theories of mixing and mass-loss in AGB stars and the formation and evolution of white dwarfs.  To study these effects, we have analyzed archival \fuse,  \hst/STIS, and Keck HIRES spectra of vZ 1128, the well-known UV-bright star in the globular cluster M3 (NGC~5272).

The star was first catalogued by \citet{von_zeipel08}.  It was studied spectroscopically by \citet{strom_strom70}, who found it to be a cluster member of late-O spectral type, with an effective temperature 31,500 K $<$ \teff\ $<$ 35,000 K, a surface gravity 3.9 $<$ \logg\ $<$ 5.2, a stellar mass $M > 0.6$ \msun, and a helium content similar to that of normal Population~I stars.  The spectrum showed absorption from N, O, and Si, but its low resolution precluded a detailed abundance analysis.  By comparing 11 UV-bright stars in globular clusters with post-HB evolutionary tracks, \citet{strom_etal70} concluded that most evolved from HB stars, while the three brightest (including vZ~1128) are post-AGB objects.  \citet{garrison_albert86} derived a spectral type of O8p.

The star's high temperature makes it a perfect target for far-ultraviolet spectroscopy.  It was observed with the {\em International Ultraviolet Observer}\/ by \citet{de_boer85}, who derived a temperature \teff\ = $30,000 \pm 2000$ K, a surface gravity $\log g = 4.0$, and a luminosity $\log L/L_{\sun} = 3.10$.  \citet{buzzoni_etal92} pointed out that these parameters place the star on the post-AGB evolutionary tracks of \citet{schoenberner83} and concluded that the star is a {\em bona fide}\/ post-AGB object.  \citet{dixon_davidsen_ferguson94} observed vZ~1128 with the Hopkins Ultraviolet Telescope and derived \teff\ = $35,000 \pm 1000$ K and $\log g = 4.0 \pm 0.25$.

Because vZ~1128 ($l = 42.5, b = +78.8$) lies 10 kpc above the Galactic plane along a line of sight with virtually no extinction ($E(B-V) = 0.01$; \citealt{harris_96}, 2010 edition), it is often used as a probe of interstellar gas in the Galactic halo \citep[\eg,][]{de_boer_savage84}.  Both our \fuse\/ \citep{howk_sembach_savage03} and STIS \citep{howk_sembach_savage06} data were originally obtained to support studies of the Galactic halo.  In \S~\ref{sec_observations} we describe these observations and our reduction of the data.  We present our determination of the star's atmospheric parameters in \S~\ref{atmospheric_parameters} and our abundance analysis in \S~\ref{abundance}.  We model the star's wind features in \S~\ref{mass_loss}.  We discuss our results in \S~\ref{sec_discussion} and summarize our conclusions in \S~\ref{sec_summary}.

\section{Spectroscopic Observations}\label{sec_observations}

\subsection{Keck Spectroscopy}

vZ~1128 was observed using the HIRES echelle spectrograph on the Keck~I telescope (Table~\ref{tab:log_obs}).  The spectrograph was configured to use the red cross-disperser and a slit with a width of 0\farcs861 and a length of 7\farcs0. We retrieved the extracted data from the Keck Observatory Archive (KOA). The standard KOA extraction provides one-dimensional spectra that are flat-fielded, bias and background subtracted, and wavelength calibrated. The extracted spectra consist of the relative flux as a function of wavelength and the uncertainty in the flux. The spectrum ranges from 4288 to to 6630 \AA\ and is divided into 30 spectral orders that cover 70 \AA\ on average. For wavelengths above 5200 \AA, there are gaps between the spectral orders that increase in size from a few \AA ngstroms to about 25 \AA\ as the wavelength increases. The \ion{H}{I}, \ion{He}{I}, and \ion{He}{II} lines are well contained within one order and do not spread across adjacent orders. The \ion{He}{I} $\lambda$5875 and \ion{He}{II} $\lambda$4686 lines are not used in the spectral analysis, because the \ion{He}{I} $\lambda$5875 line falls in a gap between the spectral orders 23 and 24, and a defect is present in the red wing of the \ion{He}{II} $\lambda$4686 line. 

\begin{table*}
 \centering
 \begin{minipage}{140mm}
\caption{Summary of the Keck, {\it FUSE}, and STIS observations.
\label{tab:log_obs}}
    \begin{tabular}{@{}lcccccc@{}}
    \hline
Data ID & Date & Grating & $R\equiv\lambda/\Delta\lambda$ & Wavelength (\AA) & Exp. Time (s) & Instrument \\
   \hline
HI.19960604.30960 & Jun 4 1996 & $\cdots$ & 47,800 & 4300--6700 & 3000 & HIRES \\
P1014101 & Jun 18 2000 & $\cdots$ &20,000& 905--1187 & 9151 & {\it FUSE} \\
P1014102 & Jun 19 2000 & $\cdots$ &20,000& 905--1187 & 8248 & {\it FUSE} \\
P1014103 & Jun 22 2000 & $\cdots$ &20,000& 905--1187 & 15610 & {\it FUSE} \\
O6F502010 & Aug 17 2002 & E140M & 45,000 & 1140--1735 & 1960 & STIS \\
O6F502020 & Aug 17 2002 & E140M & 45,000 & 1140--1735 & 10712 & STIS \\
O6F501010 & Aug 16 2002 & E230M & 30,000 & 1574--2382 & 1930 & STIS \\
O6F501020 & Aug 16 2002 & E230M & 30,000 & 1574--2382 & 8034 & STIS \\
O6F501030 & Aug 16 2002 & E230M & 30,000 & 2303--3111 & 2784 & STIS \\
\hline
\end{tabular}
\end{minipage}
\end{table*}

\begin{figure}
\includegraphics[width=87mm]{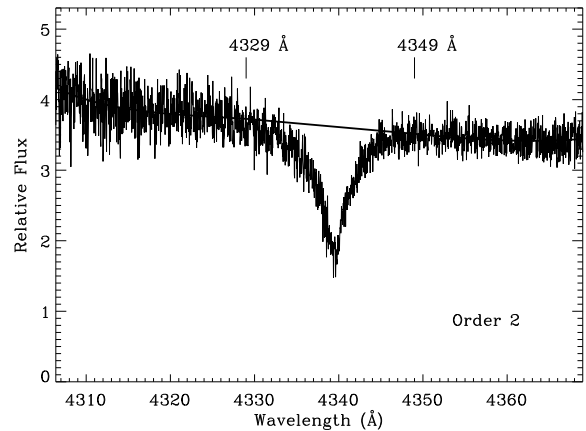}
\caption{vZ~1128's Keck HIRES spectral order number 2 showing H$\gamma$. The thick curve is the best fit of the continuum.   
\label{fig:hires_gamma_order}}

\end{figure}

We normalized the individual spectral orders by fitting the continuum with fifth-order polynomials and dividing the continuum by the  fits. We paid special attention to the spectral orders that contain the Balmer lines H$\alpha$, H$\beta$, and H$\gamma$, because the lines span significant portions of these orders. We fitted the continuum on both sides of the Balmer lines with a fifth-order polynomials, then drew a straight line using only the end-points of the polynomials. The straight lines span the Balmer lines. The end points are 6551 and 6571 \AA\ for H$\alpha$, 4851 and 4871 \AA\ for H$\beta$, and 4329 and 4349 \AA\ for H$\gamma$.  Figure~\ref{fig:hires_gamma_order} shows such a fit for H$\gamma$. The normalization is carried out by dividing the continuum by the best-fit polynomials and by the straight lines that cover the Balmer lines.  Finally, the spectral orders were merged to form a single normalized spectrum. 



Table~\ref{tab:hires_lines} gives the equivalent widths of the Balmer and helium lines, as well as of the \ion{N}{III} and \ion{O}{III} lines that are observed in the HIRES spectrum. Our equivalent widths are significantly smaller than those measured by \cite{strom_strom70} from a Carnegie image-tube spectrogram of vZ~1128 obtained with the Hale telescope at the Palomar Observatory. The resolution of their spectrogram was roughly 1--2 \AA. The equivalent widths that they measured for the \ion{He}{I} $\lambda$4388, \ion{He}{I}  $\lambda$4471, and  \ion{He}{II} $\lambda$4542 line are 560~m\AA, 1000~m\AA, and 1050~m\AA, respectively. Table~\ref{tab:hires_lines} shows that our measurements give equivalent widths that are smaller by 42\%, 34\%, and 63\%. \cite{strom_strom70} also measured the equivalent widths of elements other than He. They tentatively identified a few metal lines that they attributed to N, O, and Si, but given the resolution of their observation, they acknowledged the possibility of blends.   The HIRES spectrum shows that only a handful of N and O lines are detected. The \ion{N}{III} $\lambda$4511 and $\lambda$4515 lines, and the \ion{O}{III} $\lambda$5592 line are in absorption, while the \ion{N}{III} $\lambda$4634, $\lambda$4641 and $\lambda$4867 lines are in emission. A handful of faint features remain unidentified. It is difficult to know whether they are real or not.

\begin{table}
  \caption{Photospheric lines observed in the HIRES spectrum of vZ~1128. \label{tab:hires_lines}}
\begin{tabular}{@{}lrrrr@{}}
\hline
Ion & $\lambda_{\rm{lab}}$ & $\log gf$ & $E_l$ & E.W. \\
     &  (\AA) & & (cm$^{-1}$) & (m\AA) \\
\hline
H$\gamma$ & 4340.46  & $-0.447$ & 82259. & $2752\pm50$  \\
H$\beta$   & 4861.32  & $-0.020$ & 82259. & $2680\pm42$  \\
H$\alpha$ & 6562.80  & $0.710$  & 82259. & $2753\pm41$$^{\rm{a}}$  \\
\ion{He}{I} & 4387.929 & $-0.880$ & 193918.391 & $326.5\pm12.1$  \\
            & 4437.551 & $-2.030$ & 171135.000 & $42.8\pm6.4$  \\
            & 4471.5   & $0.052$ & 169087.     & $664.6\pm15.3$  \\
            & 4713.2   & $-0.973$ & 169087.     & $215.4\pm7.3$  \\
            & 4921.931 & $-0.430$ & 171135.000  & $367.3\pm13.4$  \\
            & 5015.678 & $-0.820$ & 166277.547  & $209.2\pm6.4$  \\
            & 5047.738 & $-1.600$ & 171135.000  & $81.0\pm7.1$  \\
\ion{He}{II} & 4541.593 & $-0.223$ & 411477.925  & $389.3\pm21.2$  \\
         & 5411.516 & $0.321$  & 411477.925  & $524.5\pm19.8$  \\
         & 6560.088$^{\rm{b}}$ & $0.759$  & 421353.135  & $\cdots$  \\
\ion{N}{III}  & 4510.96  & $-0.059$ & 287591.5    & $11.8\pm2.8$  \\
            & 4514.85  & $0.221$  & 287706.9    & $24.3\pm4.4$  \\
& 4634.13  & $-0.086$ & 245665.4    & $-30.6\pm4.3$$^{\rm{c}}$ \\
& 4640.64  & $0.168$  & 245701.3    & $-43.9\pm4.6$$^{\rm{c}}$ \\
& 4867.17 & $0.341$ & 309849.8 & $-12.4\pm3.5$$^{\rm{c}}$ \\ 
\ion{O}{III}  & 5592.252 & $-0.337$ & 273081.33   & $48.4\pm4.4$  \\
\hline
\end{tabular}
{$^{\rm{a}}$E.W. includes that of the \ion{He}{II} line at 6560.088 \AA.}\\
{$^{\rm{b}}$Line blended with blue wing of H$\alpha$.}\\
{$^{\rm{c}}$Line in emission.}
\end{table}

\subsection{\fuse\/ Spectroscopy}\label{sec_fuse_data}

The {\em Far Ultraviolet Spectroscopic Explorer (FUSE)}\/ consists of four independent spectrographs, two with
LiF-coated optics that are sensitive to wavelengths between 990 and 1187 \AA, and two with SiC-coated optics that provide reflectivity to wavelengths below the Lyman limit.  The four channels overlap between 990 and 1070~\AA.  For a complete description of \fuse, see \citet{moos00} and \citet{sahnow00}.  Observations of vZ~1128 totaling 31.5 ks were obtained through the \fuse\/ $30\arcsec \times 30\arcsec$ aperture in 2000 June \citep{howk_sembach_savage03}.  A summary of the \fuse\/ observations is presented in Table~\ref{tab:log_obs}.  For this project, the data were reduced using v3.1.3 of CalFUSE, the standard data-reduction pipeline software.  (A more recent version of the program, v3.2.0, is described by \citealt{dixon_etal07}.)  CalFUSE corrects for a variety of instrumental effects, extracts spectra from each of the four \fuse\/ channels, and performs wavelength and flux
calibration.  The extracted spectra are binned by 0.013 \AA. The spectra from each exposure were aligned by cross-correlating on the positions of stellar absorption features and combined into a single spectrum for each channel.  

In their analysis of the {\it FUSE} spectrum of vZ~1128, \citet{howk_sembach_savage03} determined the radial velocity of the star with respect to the local standard of rest (LSR). Because the zero point of the wavelength scale of individual {\it FUSE}\/  spectra cannot be determined with accuracy, they compared the \ion{H}{I} 21~cm emission line profile from \citet{danly_etal92} to the absorption profiles corresponding to the \ion{Ar}{I} lines at 1048.220 \AA\ and 1066.660 \AA. This allowed them to shift the {\it FUSE}\/ LiF1A spectrum and the other channels to the LSR. They based their comparison on the assumption that the \ion{H}{I} emission, when converted to an absorption-line profile, is similar to the Ar profiles. They showed that the neutral and ionized gas along the line of sight of vZ~1128 has LSR velocities between $-70\ {\rm{km\, s}}^{-1}$ and $+30\ {\rm{km\, s}}^{-1}$. After shifting the {\it FUSE}\/ spectra to the LSR, Howk et al.\ measured the velocity of about 10 stellar lines, obtaining an average velocity of $V_{\rm{LSR}} = -140\pm8\ {\rm{km\, s}}^{-1}$. The heliocentric radial velocity is then $V_{\rm{R}} = V_{\rm{LSR}} - V_{\sun} \cos 57\degr  = -150.9$ km$\,$s$^{-1}$, where $V_{\sun} = + 20$ km$\,$s$^{-1}$ is the solar velocity, and $57\degr$ is the angle on the celestial sphere between the Sun's apex and vZ~1128. This radial velocity is  close to the value $V_{\rm{R}} = -153\pm15\ {\rm{km\, s}}^{-1}$ measured by \citet{strom_strom70}.

\begin{figure*}
\includegraphics[width=180mm]{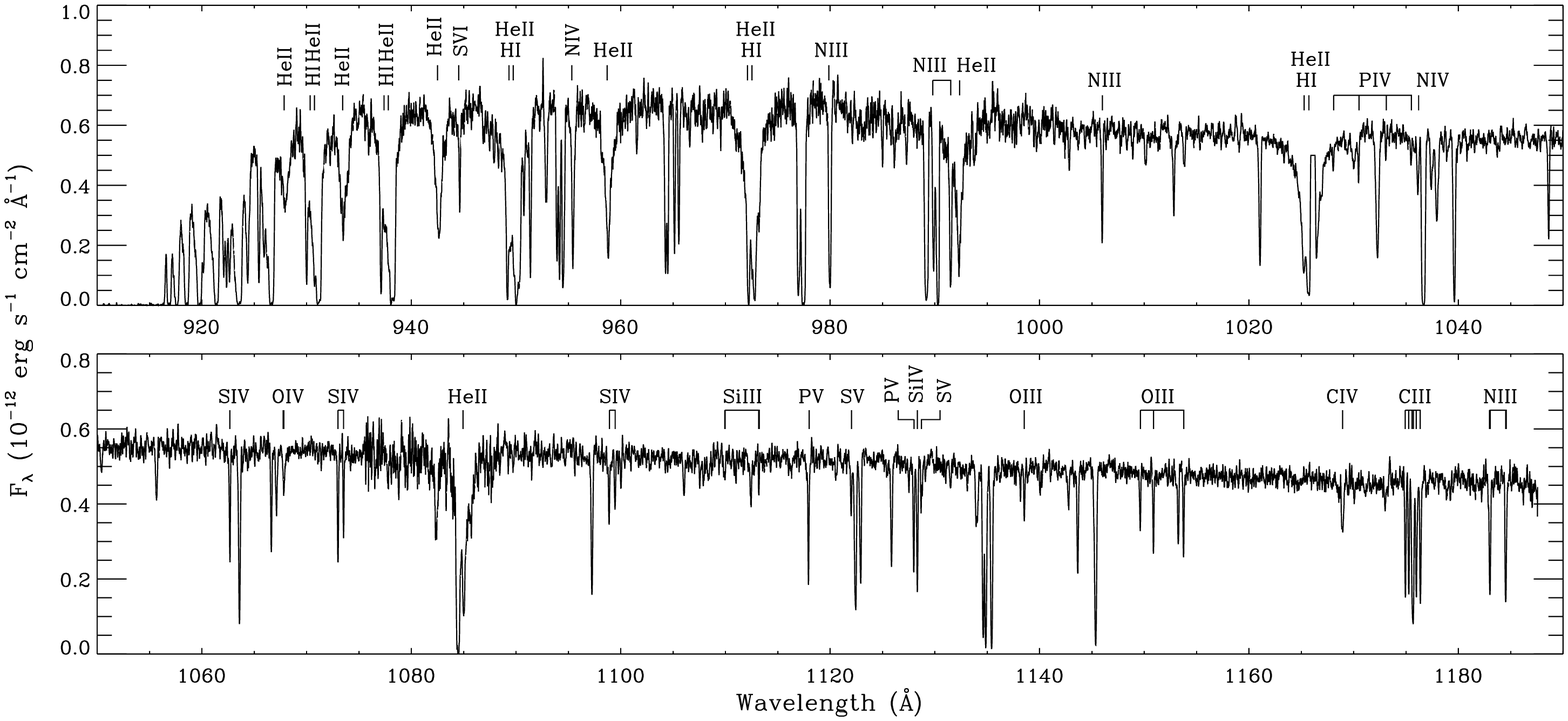}
\caption{{\it FUSE} spectrum of vZ~1128.  This spectrum was obtained by merging the segments SiC1B, LiF1A, SiC2B, LiF2A, and LiF1B.  The strongest photospheric lines are identified above the spectrum.  Most unmarked absorption lines are interstellar lines. Their identification is presented in Fig.\ 2 of \citet{howk_sembach_savage03}, who analyzed the line of sight toward vZ~1128.  \label{fig:fuse_spec}}
\end{figure*}

\citet{howk_sembach_savage03} used the \fuse\ spectrum of vZ~1128 to study the interstellar matter along the line of sight to the star. They identified about two dozen interstellar lines from ions such as \ion{C}{II}, \ion{C}{III}, \ion{N}{I}, \ion{O}{I}, \ion{O}{VI}, \ion{Si}{II}, \ion{P}{II}, \ion{S}{III}, \ion{Ar}{I}, \ion{Fe}{II}, and \ion{Fe}{III}. \fig{fig:fuse_spec} shows the \fuse\ spectrum of vZ~1128 and identifies the strongest stellar lines. Their properties are summarized in Table~\ref{tab:atomic_data}.  The stellar features are easily identifiable, because their radial velocities are well separated from those of the interstellar lines. Also, with an effective temperature of 36,600~K, the dominant ions in the photosphere of vZ~1128 are two, three, or four times ionized. Surprisingly, the \fuse\/ spectrum of vZ~1128 does not show numerous stellar lines. About three dozen lines corresponding to elements such as He, C, N, O, Si, P, and S are detected. No lines from Fe, iron peak elements, or elements beyond Zn are detected. The equivalent widths of the stellar lines (other than the H and He lines) range from $\sim10$~m\AA\  to $\sim170$~m\AA. The strongest of these features are the \ion{C}{III} $\lambda$1176; the \ion{N}{III} $\lambda$980, $\lambda$991, $\lambda$1006, and $\lambda$1183; the \ion{N}{IV} $\lambda$923 and $\lambda$955; and the \ion{S}{IV} $\lambda$1068 multiplets. 

%
\begin{table*}
\centering
 \begin{minipage}{110mm}
\caption{Photospheric lines used for measuring the abundances in vZ~1128. \label{tab:atomic_data}}
\begin{tabular}{@{}lrrrrrc@{}}
\hline
Ion & $\lambda_{\rm{lab}}$ & $\log gf$ & $E_l$ & E.W. & $\log N({\rm{X}})/N({\rm{H}})$ & Instrument \\
 & (\AA) &  & (cm$^{-1}$) & (m\AA) & & \\
\hline
\ion{C}{III}  & 1175.665$^{\rm{a}}$ & 0.389 & 52419.400 & $594.0\pm14.0$ & $-5.49\pm0.09$ & \fuse \\
            & 1175.665$^{\rm{a}}$ & 0.389 & 52419.400 & $607.0\pm14.6$ & $-5.42\pm0.12$ & STIS \\
            & 1247.383 & $-0.314$ & 102352.040 & $65.8\pm4.3$ & $-5.79\pm0.09$ & STIS \\
            & 1296.377$^{\rm{a}}$ & $0.537$  & 270013.000 & $9.4\pm2.4$& $-5.66\pm0.18$ & STIS \\
            & 2297.578 & $-0.264$ & 102352.040 & $116.5\pm6.5$ &$-5.51\pm0.17$ & STIS \\
\ion{C}{IV}  & 1168.933$^{\rm{a}}$ & $ 0.640$ & 324886.094 & $69.9\pm4.8$ & $-5.72\pm0.13$ & \fuse \\
& 1168.933$^{\rm{a}}$ & $ 0.640$ & 324886.094 & $81.9\pm6.9$ & $-5.65\pm0.15$ & STIS \\
            & 1548.195 & $-0.419$ & 0.000 & $282.4\pm10.0$ & $-5.48\pm0.04$ & STIS \\
            & 1550.772 & $-0.720$ & 0.000 & $227.9\pm8.5$ & & STIS \\
\ion{N}{III}  &  979.876$^{\rm{a}}$ & 0.153    &  101027.00 & $282.3\pm8.2$ & $-4.42\pm0.13$ & \fuse \\
            & 989.799  & $-0.610$ & 0.000      & $171.5\pm8.2$ & $-4.42\pm0.09$ & \fuse  \\
            & 991.511  & $-1.317$ & 174.400  & $242.0\pm10.4$& & \fuse \\
            & 991.577  & $-0.357$ & 174.400  &               & & \fuse \\
            & 1005.993 & $-0.807$ & 131004.300 & $100.8\pm6.2$ & $-4.63\pm0.09$ & \fuse \\
            & 1006.036 & $-1.123$ & 131004.300 &            &  & \fuse \\
            & 1182.971 & $-0.924$ & 145875.700 & $113.2\pm5.9$ & $-4.58\pm0.12$ & \fuse \\
            & 1183.032 & $-0.608$ & 145875.700 &          & &\fuse   \\
            & 1184.514 & $-0.212$ & 145985.800 & $111.7\pm5.4$ & & \fuse \\
            & 1184.574 & $-0.915$ & 145985.800 &    & & \fuse \\
            & 1182.971 & $-0.924$ & 145875.700 & $105.5\pm4.4$ & $-4.45\pm0.11$ & STIS \\
            & 1183.032 & $-0.608$ & 145875.700 &          & & STIS   \\
            & 1184.514 & $-0.212$ & 145985.800 & $115.0\pm4.9$ & & STIS \\
            & 1184.574 & $-0.915$ & 145985.800 &    & & STIS \\
      & 4510.96  & $-0.059$ & 287591.500 & $11.8\pm2.8$ & $-4.49\pm0.11$ & HIRES \\
      & 4514.85  & $0.221$  & 287706.900 & $24.3\pm4.4$ & & HIRES \\
      & 4634.13 & $-0.086$ & 245665.4 & $-30.6\pm4.3^{\rm{b}}$ & $-4.39\pm0.15$ & HIRES \\
      & 4640.64 & $0.168$ & 245701.3  & $-43.9\pm4.6^{\rm{b}}$ & & HIRES \\
      & 4867.17 & 0.341 & 309849.8 & $-12.4\pm3.5^{\rm{b}}$ & $-4.20\pm0.10$ & HIRES \\
\ion{N}{IV}  & 921.994  & $-0.551$ &  67272.300 & $127.9\pm10.6$& $-4.38\pm0.04$ & \fuse \\
            & 922.519  & $-0.648$ &  67209.200 & $162.4\pm10.7$&              &  \fuse \\
            & 924.284  & $-0.552$ &  67416.300 & $164.5\pm9.4$ &  &\fuse   \\
            & 955.334  & $-0.399$ & 130693.900 & $213\pm10.1$ & $-4.35\pm0.06$ & \fuse \\
            & 1036.119 & $0.778$ & 420058.000 & $72.5\pm4.6$ & $-4.53\pm0.09$ &  \fuse \\
            & 1036.149 & $0.618$ & 420049.594 &   &   & \fuse \\
            & 1036.196 & $0.447$ & 420045.813 &   &   & \fuse \\
            & 1036.237 & $-0.285$ & 420049.594 &  &   & \fuse \\
            & 1036.239 & $-0.285$ & 420058.000 &  &   & \fuse \\
            & 1036.327 & $-1.829$ & 420058.000 &  &   & \fuse \\
            & 1078.711 & $0.599$ & 429159.600 & $22.4\pm3.5$ & $-4.46\pm0.17$ & \fuse \\
            & 1718.550 & $-0.289$ & 130693.900 & $197.0\pm21.9$ & $-4.43\pm0.22$ & STIS \\
\ion{N}{V} & 1238.821 & $-0.505$ & 0.000 & $81.9\pm4.5$ & $-6.34\pm0.10$ & STIS \\
           & 1242.804 & $-0.807$ & 0.000 & $72.8\pm4.1$ &  & STIS \\
\ion{O}{III}  & 1138.535 & $-0.755$ & 210461.797 & $35.8\pm4.0$ & $-4.55\pm0.11$ & \fuse \\
            & 1149.634 & $-1.080$ & 197087.703 & $37.9\pm3.4$ & $-4.37\pm0.09$ & \fuse \\
            & 1150.884 & $-0.603$ & 197087.703 & $52.9\pm3.7$ & & \fuse \\
            & 1153.775 & $-0.382$ & 197087.703 & $59.7\pm4.3$ & & \fuse \\
            & 5592.252 & $-0.337$ & 273081.33  & $48.4\pm4.4$ & $-4.62\pm0.12$ & HIRES \\
\ion{O}{IV}  & 1067.768 &  $0.504$ & 419533.906 & $41.0\pm4.3$ & $-4.27\pm0.14$ & \fuse \\
            & 1067.832 &  $0.658$ & 419550.594 &          &   & \fuse \\
            & 1338.615 & $-0.632$ & 180480.797 & $68.6\pm3.4$ & $-4.79\pm0.13$ & STIS \\
            & 1342.990 & $-1.333$ & 180724.203 & $46.5\pm3.2$ & & STIS  \\
            & 1343.514 & $-0.380$ & 180724.203 & $85.0\pm3.9$ & & STIS \\
\ion{O}{V}  & 1371.296 & $-0.328$ & 158797.703 & $37.5\pm3.3$ & $-4.58\pm0.16$ & STIS \\
\ion{Al}{III} & 1854.716 & $0.060$  &      0.000 & $55.3\pm8.3$ & $-6.87\pm0.14$ & STIS \\
            & 1862.790 & $-0.240$ &      0.000 & $49.2\pm12.8$ &  & STIS  \\
\ion{Si}{III} & 1109.940 & $-0.186$ & 52853.281  & $9.9\pm2.2$  & $-5.84\pm0.09$ & \fuse \\
            & 1109.970 & $0.294$  & 52853.281  &  &  & \fuse \\
            & 1113.174 & $-1.356$ & 53115.012  & $24.1\pm3.2$ & & \fuse \\
            & 1113.204 & $-0.186$ & 53115.012  &  &  & \fuse \\
            & 1113.230 & $0.564$  & 53115.012  &  &  & \fuse \\
            & 1298.946 & $0.443$ & 53115.012 & $13.9\pm1.8$ & $-6.17\pm0.06$ & STIS \\
\hline
\end{tabular}
\end{minipage}
\end{table*}

%
\begin{table*}
\centering
 \begin{minipage}{110mm}
\contcaption{Photospheric lines used for measuring the abundances in vZ~1128.}
\begin{tabular}{@{}lrrrrrc@{}}
\hline
Ion & $\lambda_{\rm{lab}}$ & $\log gf$ & $E_l$ & E.W. & $\log N({\rm{X}})/N({\rm{H}})$ & Instrument \\
 & (\AA) &  & (cm$^{-1}$) & (m\AA) & & \\
\hline
\ion{Si}{IV} & 1066.629$^{\rm{a}}$ & 0.961 & 160375.000 & $67.7\pm4.8$ & $-6.22\pm0.11$ & \fuse \\
            & 1128.325 & $-0.480$ & 71748.641 & $94.8\pm4.0$ & $-6.21\pm0.08$ & \fuse \\
            & 1128.340 & $0.470$  & 71748.641 &  & & \fuse \\
            & 1393.755 & 0.030 & 0.000 & $184.6\pm6.0$ & $-5.73\pm0.05$ & STIS \\ 
            & 1402.770 & $-0.280$ & 0.000 & $146.1\pm5.6$ & & STIS \\
\ion{P}{IV}  & 950.657  & $0.270$  &     0.000 & $46.7\pm6.3$ & $-8.18\pm0.26$ & \fuse \\
            & 1028.094 & $-0.317$ & 67918.031 &  $18.7\pm3.5$ & $-7.90\pm0.10$ & \fuse \\
 & 1030.514 & $-0.444$ & 68146.477 &  $37.2\pm3.5$ &   & \fuse \\
            & 1030.515 & $0.255$  & 68615.172 &  & & \fuse \\
            & 1033.112 & $-0.319$ & 68146.477 & $18.3\pm3.5$ &  & \fuse \\
            & 1035.516 & $-0.222$ & 68615.172 & $23.1\pm3.5$ &  & \fuse \\

\ion{P}{V}  & 1117.977 & $-0.010$ &     0.000 & $82.9\pm4.1$ & $-8.52\pm0.12$ & \fuse \\
            & 1128.008 & $-0.320$ &     0.000 & $77.2\pm3.8$ & & \fuse \\
\ion{S}{IV}  & 1062.678 & $-1.089$ &     0.000 & $72.8\pm4.2$ & $-6.61\pm0.09$ & \fuse \\
            & 1072.996 & $-0.829$ &   951.100 & $70.8\pm3.5$ & & \fuse \\
            & 1073.528 & $-1.789$ &   951.100 & $45.2\pm3.1$ & & \fuse \\
            & 1098.917 & $-0.607$ & 94150.398 & $43.8\pm4.0$ & $-6.33\pm0.11$ & \fuse \\
            & 1099.472 & $-0.799$ & 94103.102 & $34.8\pm4.1$ & & \fuse \\
\ion{S}{V} & 1122.042 & $0.094$  & 234956.000 & $41.3\pm4.1$ & $-6.19\pm0.17$ & \fuse \\
            & 1128.667 & $-0.066$ & 234947.094 & $23.3\pm2.6$ & & \fuse \\
            & 1128.776 & $-0.968$ & 234956.000 & $13.5\pm2.1$ & & \fuse \\
            & 1501.760 & $-0.504$ & 127150.703 & $71.2\pm4.4$ & $-6.70\pm0.16$ & STIS \\
\ion{S}{VI}  & 944.523  & $-0.350$ &      0.0   & $70.0\pm5.3$ & $-6.06\pm0.16$ & \fuse \\
\ion{Fe}{V} & 1373.589 & $0.312$ & 187719.000 & $22.8\pm2.6$ & $-6.26\pm0.15$ & STIS \\
     & 1373.679 & $0.188$ & 187157.500 & $9.9\pm1.9$  & & STIS \\
     & 1376.337 & $0.439$ & 188395.297 & $26.8\pm2.9$ & & STIS \\
     & 1376.451 & $0.049$ & 186725.500 & $16.4\pm2.7$ & & STIS \\
     & 1378.561 & $0.430$ & 205536.406 & $19.6\pm3.0$ & & STIS \\  
      \\
\ion{Fe}{V}& 1459.769 & $0.222$ & 213534.094 & $38.3\pm4.1$ & $-6.64\pm0.16$ & STIS\\
\ion{Fe}{IV} & 1462.586 & $-0.761$ & 190811.797 & $8.2\pm2.5$ & & STIS \\
\ion{Fe}{V} & 1462.636 & $0.041$ & 186433.594 & $15.0\pm2.5$ & & STIS \\
\ion{Fe}{IV} & 1464.695 & $0.158$ & 190318.344 & $18.2\pm3.5$ & & STIS \\
\ion{Fe}{V} & 1465.380 & $-0.260$ & 187157.500 & $15.4\pm3.1$ & & STIS\\
    & 1465.401 & $0.304$ & 221305.203 &              & & STIS\\
    & 1468.998 & $0.474$ & 217122.500 & $21.7\pm3.4$ & & STIS\\
    & 1472.095 & $0.396$ & 216860.406 & $14.6\pm3.3$ & & STIS\\
    & 1472.511 & $0.312$ & 216779.094 & $10.1\pm3.0$ & & STIS\\
\\
\ion{Fe}{IV} & 1531.223 & $0.051$ & 158738.688 & $18.1\pm3.6$ & $-5.97\pm0.15$ & STIS\\
     & 1532.630 & $-0.129$ & 128541.852 & $19.8\pm4.7$ & & STIS \\
     & 1532.903 & $0.434$ & 168566.438 & $18.6\pm4.5$ & & STIS \\
     & 1533.267 & $0.177$ & 159010.391 & $20.6\pm4.4$ & & STIS \\
     & 1533.869 & $-0.004$ & 128191.539 & $47.1\pm5.4$: & & STIS \\
     & 1542.155 & $0.045$ & 128541.852 & $20.2\pm4.7$ & & STIS \\
     & 1542.698 & $0.360$ & 128967.672 & $37.2\pm4.6$: & & STIS \\
     & 1544.486 & $0.453$ & 168526.375 & $20.0\pm4.2$ & & STIS \\
     & 1546.404 & $-0.318$ & 127929.117 & $22.8\pm4.7$ & & STIS\\ 
     \\
\ion{Ni}{IV} & 1398.193 & $0.584$ & 110410.602 & $14.5\pm2.6$ & $-7.87\pm0.12$ & STIS \\
            & 1411.451 & $0.448$ & 111195.797 & $13.7\pm2.9$ & $-7.73\pm0.11$ & STIS \\
\ion{Ni}{V} & 1244.174 & $0.443$ & 164525.906 & $10.6\pm3.9$ & $-6.95\pm0.15$ & STIS \\
            & 1264.501 & $0.320$ & 164525.906 & $7.6\pm2.1$ & $-7.19\pm0.23$ & STIS\\
\hline
\end{tabular}

$^{\rm{a}}${Multiplet.}\\
$^{\rm{b}}${Line in emission.}
\end{minipage}
\end{table*}

\subsection{STIS Spectroscopy}\label{sec_stis_data}

High-resolution STIS echelle observations of vZ~1128 were obtained through the $0\farcs2 \times 0\farcs06$ aperture of the Space Telescope Imaging Spectrograph (STIS) in 2002 August \citep[\hst\/ Proposal 9150;][]{howk_sembach_savage06}.  Observations  O6F501020 and O6F502020 employed the E140M grating and the FUV Multi-Anode Microchannel Array (MAMA) detector, and observations O6F501010, O6F501020, and O6F501030 used the E230M grating and the NUV MAMA detector.  The STIS observations are summarized in Table~\ref{tab:log_obs}. The data were retrieved from the Mikulski Archive for Space Telescopes\footnote{\url{http://archive.stsci.edu/hst/}} and calibrated using the on-the-fly version (2.15c) of CALSTIS in 2005 June.  The design and construction of STIS are described by \citet{woodgate_etal98}, and information about its on-orbit performance is provided by \citet{kimble_etal98}. Information about CALSTIS can be found in the {\em HST STIS Data Handbook}\/ \citep{STIS_Handbook}.

The STIS spectra consist of three pass bands that cover a wavelength range from about 1140 \AA\ to 3111 \AA. We measured a stellar radial velocity of $-151.8\pm3.0$ km\ s$^{-1}$ by using a dozen stellar lines. This velocity agrees with the one measured by \citet{howk_sembach_savage03} and \citet{strom_strom70}.  Unlike the \fuse\ spectra, the STIS spectra show many stellar \ion{Fe}{IV} and \ion{Fe}{V} lines. Table~\ref{tab:atomic_data} gives the atomic properties of some of the Fe lines that are observed in the STIS E140M spectrum. These \ion{Fe}{IV} and \ion{Fe}{V} transitions start from relatively high energy levels, but have large oscillator strengths. Consequently, the lines are not very strong and have equivalent widths ranging from about 10 to 50 m\AA. A handful of faint \ion{Ni}{IV} and \ion{Ni}{V} lines are also detected. 
Their equivalent widths range from about 7 to 15 m\AA.  No other iron-peak elements nor any heavier elements are detected. On the other hand, several stellar lines coming from light elements are detected. The resonance lines of the \ion{C}{IV} $\lambda$1550 doublet, the \ion{N}{V} $\lambda$1240 doublet, the \ion{Al}{III} $\lambda$1857 doublet, the \ion{Si}{III} $\lambda$1206 line, and the \ion{Si}{IV} $\lambda$1397 doublet are observed. A handful of lines starting from energy levels above the ground state are also detected. For instance, the \ion{C}{III} $\lambda$1247 and $\lambda$2297 lines, the \ion{O}{IV} $\lambda$1340 triplet, the \ion{O}{V} $\lambda$1371 line, and the \ion{S}{V} $\lambda$1501 line are observed with equivalent widths that vary from $\sim$45~m\AA\ to 85~m\AA. Table~\ref{tab:atomic_data} summarizes the properties of the lines observed in the STIS spectra.

\begin{figure}
\includegraphics[width=87mm]{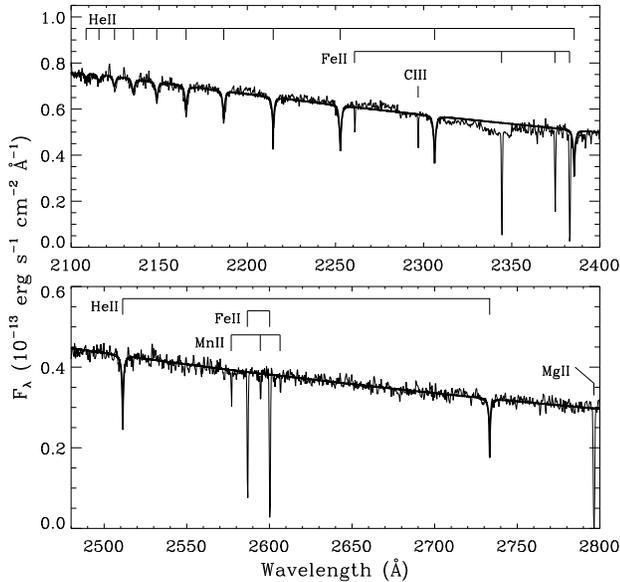}
\caption{\ion{He}{II} line series corresponding to the transitions  
$n = 3 \rightarrow 6$ up to 18 observed in the STIS 
E230M spectrum of vZ~1128. The bold curve is a 
model with $T_{\rm{eff}} = 36$,600~K, $\log g = 3.95$, and $\log 
N({\rm{He}})/N({\rm{H}}) = -0.84$ that is scaled to match the flux 
level of the STIS observation. The \ion{Mg}{II}, \ion{Mn}{II}, and 
\ion{Fe}{II} absorption features are interstellar lines, while the 
\ion{C}{III} feature is a stellar line.
\label{fig:he_series}}
\end{figure}

The STIS E230M data show the \ion{He}{II} series for transitions from the lower energy level $n = 3$ to the upper energy levels $n = 6$ through 18. In fact, because the wavelength range of the E230M O6F501020 and O6F501030 datasets covers the region between 1574 \AA\ and 3111 \AA, only the transitions $n = 3 \rightarrow 4$ and 5 are missing. These two transitions correspond to the $\lambda$4686 and $\lambda$3203 lines. \fig{fig:he_series} shows the portions of the STIS E230M spectrum where the \ion{He}{II} series is visible. The lower panel shows the \ion{He}{II}  $\lambda$2733 and $\lambda$2511 lines that correspond to the transitions $n = 3 \rightarrow 6$ and 7. The upper panel shows the remaining \ion{He}{II} lines, corresponding to the transitions  $n = 3 \rightarrow 8$ through 18. 

\section{Determination of Atmospheric Parameters}\label{atmospheric_parameters}

By comparing the Keck HIRES spectrum of vZ~1128 to a grid of stellar atmosphere models, and by using a chi-square method to find the best model, we can determine the effective temperature, gravity, and He abundance of the star.  We computed a grid of non-local thermodynamic equilibrium (non-LTE) stellar atmosphere models using the program TLUSTY \citep{hubeny_lanz95}. The models are composed solely of H and He. The atomic models for H, \ion{He}{I}, and \ion{He}{II} are similar to those that \citet{lanz_hubeny2003} used for computing their grid of O-type stars. Our grid of models covers an effective temperature range $T_{\rm{eff}} =  26$,000 to 42,000~K in steps of 2000~K, a gravity range $\log g = 3.6$ to 4.8 in steps of 0.2 dex, and a He abundance range $\log N({\rm{He}})/N({\rm{H}}) = 0.0$ to $-2.0$ in steps of 0.5 dex. From that grid of stellar atmosphere models, we computed a grid of synthetic spectra using the program SYNSPEC (I. Hubeny \& T. Lanz, private communication). The synthetic spectra cover a wavelength range of 3500 to 6800 \AA. The synthetic spectra were convolved with a Gaussian with a FWHM of 0.1 \AA\ and were normalized to replicate the Keck HIRES spectrum. We used the IDL function CURVEFIT to fit the synthetic spectra to the observed Balmer and He lines with the effective temperature, gravity, and He abundance as free parameters. 

\begin{figure}
\includegraphics[width=87mm]{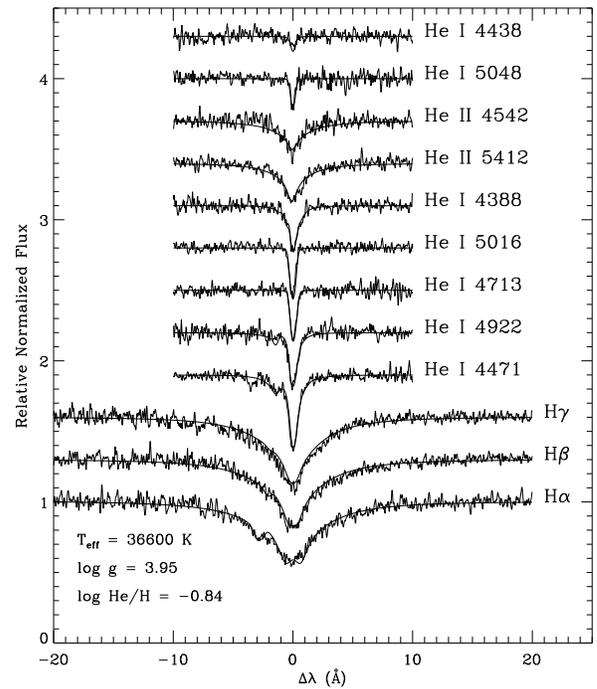}
\caption{Best-fit model plotted over the Balmer, \ion{He}{I}, and \ion{He}{II} lines observed in the Keck HIRES spectrum of vZ~1128. The faint absorption feature in the blue wing of H$\alpha$ is the \ion{He}{II} $\lambda$6560 line. The \ion{He}{I} 
$\lambda$5875 line is not detected in the HIRES spectrum, because the line is located in a gap between two spectral orders. The \ion{He}{II} $\lambda$4686 line is not shown, because the line is corrupted by an artifact and is not usable by the fitting procedure.
\label{fig:hires_spec}}

\end{figure}

\fig{fig:hires_spec} shows our best fits to the Balmer and He lines that are observed in the HIRES spectrum. Table~\ref{tab:hires_lines} gives the atomic parameters and equivalent widths of these lines. The best fit yields an effective temperature $T_{\rm{eff}} = 36$,$600\pm400$~K, a gravity $\log g = 3.95\pm0.10$, and a He abundance $\log N({\rm{He}})/N({\rm{H}}) = -0.84\pm0.03$. The uncertainties of the atmospheric parameters are from the quality of the fit and the error associated with our experiments regarding the normalization of the HIRES spectrum. By slightly changing the parameters of the normalization, we obtained slightly different atmospheric parameters. This source of error is incorporated into the uncertainties cited above. \fig{fig:hires_spec} shows that the fit is very good in general. The fit of H$\alpha$ shows some small discrepancies in the core of the line and just next to the  \ion{He}{II} $\lambda$6560 line in the blue wing of H$\alpha$. The fits of the H$\alpha$ wings and the \ion{He}{II} $\lambda$6560 line are quite good though. The model matches H$\beta$ over the whole line profile, while it shows some discrepancies close to the core of H$\gamma$. This discrepancy could indicate that the normalization of the HIRES spectrum around H$\gamma$ is not as good as the normalization of H$\alpha$ and H$\beta$.  The forbidden component  of the \ion{He}{I} $\lambda$4471 line is well reproduced by the model, but the computed forbidden component of the \ion{He}{I} $\lambda$4922 line is slightly too strong to account for the observation. The kink in the blue wing of the \ion{He}{I} $\lambda$4388 line is caused by its forbidden component. There is a feature in the red wing of the \ion{He}{II} $\lambda$5412 line that is not accounted for, and the computed line profile of the \ion{He}{II} $\lambda$4542 line seems slightly too strong. This could indicate, however, that the normalization of the observed \ion{He}{II} $\lambda$4542 line is not perfect.



\section{Abundance Analysis}\label{abundance}


\subsection{Technique}\label{technique}

The abundance of an element is measured by comparing its observed lines to  a set of synthetic spectra that are computed for different abundances.  We computed grids of non-LTE atmosphere models by adopting the atmospheric parameters derived in \S~\ref{atmospheric_parameters}, $T_{\rm{eff}} = 36$,$600\pm400$~K, $\log g = 3.95\pm0.10$, and $\log N({\rm{He}})/N({\rm{H}}) = -0.84\pm0.03$, and by considering several chemical compositions of five elements at a time. The different compositions that we considered consist of $\rm{H}+\rm{He}+\rm{N}+\rm{O}+\rm{X}$, where X is either C, Al, Si, P, S, or Fe. The models were computed with $\log N({\rm{N}})/N({\rm{H}}) = -4.4$ and $\log N({\rm{O}})/N({\rm{H}}) = -4.5$, and by considering 15 values of the abundances $\log N({\rm{X}})/N({\rm{H}}) =  -4.0$ to $-9.2$, in steps of 0.4 dex. The N and O abundances were determined by considering models with a chemical composition of H+He+N+O. These two elements were included in all model calculations because they are the most abundant metals in the stellar photosphere. 

Using the non-LTE stellar atmosphere models, we computed synthetic spectra for each abundance and for all the stellar lines observed in the {\it FUSE}, STIS, and Keck spectra. The resolution of the synthetic spectra was adjusted to match the resolution of the observed spectra. We used the programs TLUSTY and SYNSPEC to compute the non-LTE stellar atmosphere models and synthetic spectra. In order to take into account the uncertainties of the atmospheric parameters, we also computed stellar atmosphere models that include the uncertainties of the effective temperature and gravity ($\Delta T_{\rm{eff}} = 400$~K  and $\Delta \log g = 0.10$ dex), and repeated the calculations by using the new sets of atmospheric parameters: $T_{\rm{eff}} =37$,000~K and $\log g = 3.95$;  $T_{\rm{eff}} =36$,600~K and $\log g = 4.05$. 

We use a chi-square minimization method to fit each line or multiplet individually. This method determines the best match between the synthetic spectra and the observations, and estimates the abundance and its uncertainty. We first normalize both the observed and theoretical absorption lines. The observed lines are normalized by dividing the spectrum by a low-order polynomial that fits the continuum on both sides of the lines of interest. The synthetic spectra are divided by the theoretical continuum computed by the program SYNSPEC. We then shift the observed spectrum to the laboratory rest frame and select the portion of the spectrum that includes the lines of interest for a given element. We use the IDL function CURVEFIT to fit the synthetic spectra to the observed lines with the abundance as a free parameter. \fig{fig:line_broadening} (top row) shows such fits for three absorption lines observed with three different instrument setups. The \ion{Si}{IV} $\lambda$1066 line is observed with the {\it FUSE} LiF1A segment, the \ion{C}{III} $\lambda$1247 line is observed with the STIS E140M grating using the NUV MAMA detector, and the \ion{C}{III} $\lambda$2297 line is observed with the STIS E230M grating using the FUV MAMA detector.

\subsection{Additional Broadening}\label{broadening}

\begin{figure}
\includegraphics[width=87mm]{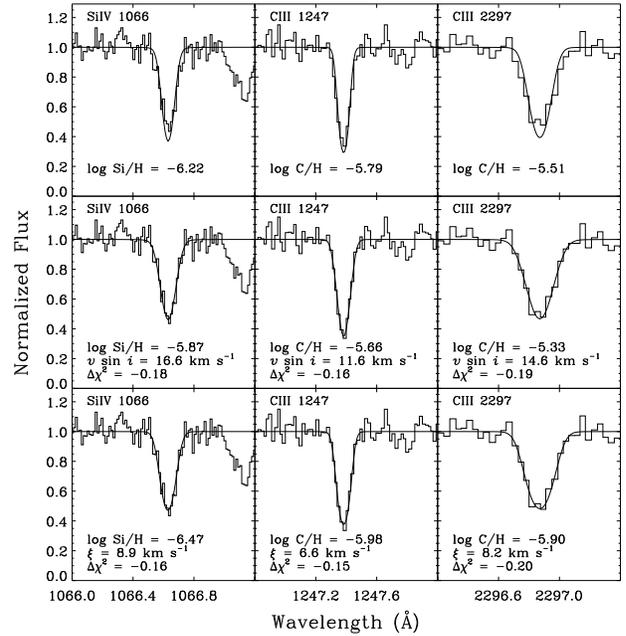}
\caption{Best fits to the \ion{Si}{IV} 
$\lambda$1066 line, the \ion{C}{III} $\lambda$1247 line, and the 
\ion{C}{III} $\lambda$2297 line when considering {\it i)} the abundance 
as the only free parameter ({\it top row}); {\it ii)} the abundance and the projected rotational velocity $v \sin i$ as free parameters
({\it middle row}); and {\it iii)} the abundance and the microturbulent velocity $\xi$ as free parameters ({\it bottom row}).
\label{fig:line_broadening}}
\end{figure}

At first sight, the match between the synthetic spectra and the observed lines plotted in \fig{fig:line_broadening} (top row) seems quite good, but the cores of the models are systematically too deep and the wings are too narrow. A better match between the observed data and model is obtained by considering an additional source of broadening, such as the projected stellar rotational velocity $v \sin i$ or the microturbulent velocity $\xi$. For example, \fig{fig:line_broadening} (middle row) shows the best fits to the same lines, but considering both the abundance and the rotational velocity as free parameters. By eye we can see that the quality of the fits has improved. Though the changes are small, the cores and wings of the lines better match the observations. 
The $\Delta\chi^2$ values displayed in \fig{fig:line_broadening} (middle row) indicate that  $\chi^2$ is significantly lower when we vary both the abundance and rotational velocity than when we vary the abundance alone. 
The average value of $v \sin i$ for these three measurements is 14.3~km~s$^{-1}$. The better fits accommodate slightly higher abundances.

Microturbulence is another possible source of line broadening. The microturbulent velocity $\xi$ is usually introduced to remove the variation of the abundance as a function of the equivalent width for lines of a given ion.  Unfortunately, given the small number of absorption lines observed in the FUV and visible spectra of vZ~1128, we cannot independently determine the microturbulent velocity $\xi$ by using this method. We can estimate the effect of microturbulence on the shape of the metal lines, though, and see whether it is a plausible explanation for the additional broadening. \fig{fig:line_broadening} (bottom row) shows our best fits when considering the abundance and the microturbulent velocity as free parameters. The quality of the fits is as good as those for which we considered the abundance and the projected rotational velocity as free parameters.  This is illustrated by the negative $\Delta\chi^2$ values, which indicate that these fits yield lower $\chi^2$ than the fits illustrated in the top row of \fig{fig:line_broadening}. The average value of $\xi$ for these three measurements is 7.9~km~s$^{-1}$. Unlike rotation, the inclusion of microturbulence in the model decreases the derived abundance. The lines in question are saturated, so increasing $\xi$ broadens them, reducing the saturation and increasing the absorption for a fixed abundance.

Based on spectroscopic analysis alone, we face the dilemma of choosing the source of the line broadening. As discussed above, we cannot measure independently the microturbulent velocity by comparing the equivalent widths and abundances of absorption lines for a given ion. Neither do we have independent information on the projected rotational velocity of the star. Given that both sources of line broadening yield practically identical results, we cannot rule out one or the other solely on the basis of the spectral observations. Both sources of line broadening could be present simultaneously in different degrees. Consequently, instead of choosing one particular source of line broadening, we have adopted the approach of considering 1) the abundance as the only free parameter; 2) the abundance and the stellar projected rotational velocity $v \sin i$ as free parameters; and 3) the abundance and the microturbulent velocity $\xi$ as free parameters. This approach allows us to move forward with our abundance analysis. It also shows us the effects of including either the stellar rotation or the microturbulent velocity on the determination of the abundances for all species observed in the atmosphere of vZ~1128. 

%
\begin{table*}
\centering
\begin{minipage}{160mm}
\caption{Comparison between the abundances observed in vZ~1128 and those in HB and RGB stars in M~3. \label{tab:abundances}}
\begin{tabular}{@{}lccccccccc@{}}
\hline
 & \multicolumn{6}{c}{vZ 1128} & &\multicolumn{2}{c}{M 3}  \\
 \cline{2-7}\cline{9-10}
 &  & \multicolumn{2}{c}{Stellar Rotation} & & \multicolumn{2}{c}{Microturbulence} & & &\\
\cline{3-4} \cline{6-7} \\
Elements & $\log N({\rm{X}})/N({\rm{H}})$ & $\log N({\rm{X}})/N({\rm{H}})$ & $v \sin i$ & & $\log N({\rm{X}})/N({\rm{H}})$ & $\xi$ & & $\log N({\rm{X}})/N({\rm{H}})$ & Ref. \\
 & & & (km s$^{-1}$) & & & (km s$^{-1}$) & & & \\
\hline
He& $-0.84\pm0.03$ & $\cdots$ & $\cdots$ & & $\cdots$ & $\cdots$ & & $-1.50\pm0.39$ & 1 \\
C & $-5.59\pm0.13$ & $-5.43\pm0.15$ & $13.1\pm1.4$ & & $-6.13\pm0.23$ & $7.8\pm1.0$ & & $-5.79\pm0.15$ & 2 \\
N & $-4.44\pm0.15$ & $-4.45\pm0.11$ & $15.6\pm3.4$ & & $-4.73\pm0.15$ & $6.2\pm2.7$ & & $-4.69\pm0.28$ & 2 \\
O & $-4.53\pm0.19$ & $-4.41\pm0.23$ & $14.1\pm1.4$ & & $-4.85\pm0.42$ & $6.9\pm0.4$ & & $-4.37\pm0.15$ & 3 \\
Al & $-6.87\pm0.14$ & $-6.54\pm0.14$ & $14.7\pm3.1$ & & $-6.99\pm0.14$ & $8.8\pm2.3$  & & $-6.60\pm0.36$ & 3 \\
Si & $-6.01\pm0.22$ & $-5.92\pm0.09$ & $15.4\pm1.6$ & & $-6.75\pm0.40$ & $8.3\pm1.0$ & & $-5.73\pm0.06$  & 3 \\
P & $-8.20\pm0.31$ & $-8.18\pm0.12$ & $13.4\pm1.1$ & & $-9.50\pm0.12$ & $7.3\pm0.8$ & & $\cdots$ & $\cdots$ \\
S & $-6.38\pm0.27$ & $-5.97\pm0.16$ & $17.3\pm1.5$ & & $-6.96\pm0.38$ & $9.5\pm3.2$ & & $\cdots$ & $\cdots$ \\
Fe & $-6.29\pm0.34$ & $-5.91\pm0.19$ & $12.7\pm1.5$ & & $-6.29\pm0.32$ & $8.4\pm0.9$ & & $-6.06\pm0.06$ & 3 \\
Ni & $-7.43\pm0.44$ & $\cdots$ & $\cdots$ & & $\cdots$ & $\cdots$ & & $-7.38\pm0.04$ & 3 \\
\hline
\end{tabular}
{REFERENCES. --- (1) \citet{behr_etal03}; (2) \citet{smith_etal96}; (3) \citet{sneden_etal04}.}
\end{minipage}
\end{table*}

%
%
%
%
%

\subsection{Results}

Table~\ref{tab:atomic_data} lists the atomic and observed properties of the stellar lines that we selected to measure the abundances. The abundances quoted in the table were measured by assuming no additional line broadening. The table shows the abundances for individual transitions and ions. The uncertainties of the abundances include the quality of the fit, the oscillator strength uncertainty, and the atmospheric parameter uncertainties ($\Delta T_{\rm{eff}}$  and $\Delta \log g$). The oscillator strength uncertainties are based on the National Institute of Standard and Technology website.\footnote{\url{http://physics.nist.gov/PhysRefData/ASD/lines\_form.html}} The different contributions to the uncertainties are combined in quadrature. Table~\ref{tab:abundances} presents the results of our abundance analysis. The second column gives the abundances that are measured without any additional broadening; the third and forth columns give the parameters that are measured when both the abundance and projected rotational velocity are considered; the fifth and sixth columns give the parameters that are measured when both the abundance and microturbulent velocity are considered. The abundances and velocities are the averages of individual transitions. The uncertainties are the standard deviation of the measurements. Before discussing the result of our abundance analysis, we will report on our analysis of the individual elements.

\subsubsection{Helium}

As reported earlier, the best fit to the HIRES spectrum yields a He abundance of $\log N({\rm{He}})/N({\rm{H}}) = -0.84\pm0.03$.  \fig{fig:hires_spec} shows the \ion{He}{I} and \ion{He}{II} lines that were included in the fit.  This He abundance is consistent with the strength of the \ion{He}{II} line series $n = 3 \rightarrow 6$ through 18 that is observed in the STIS spectrum (\fig{fig:he_series}). It is also consistent with the strength of the \ion{He}{II} line series $n = 2 \rightarrow 3$ through 15 that is observed in the {\em FUSE}\/ and STIS spectra. 

\subsubsection{Carbon}

\ion{C}{III} and \ion{C}{IV} lines are observed in both {\it FUSE} and STIS spectra. No \ion{C}{II} lines are detected. The resonance \ion{C}{III} line at 977.020 \AA\ is not listed in Table~\ref{tab:atomic_data}, because it is blended with the strong interstellar medium absorptions of \ion{O}{I} and \ion{C}{III}. The \ion{C}{IV} lines at 1548.195 and 1550.772 \AA\ are the strongest carbon lines. Although the red wings of the stellar doublet are blended with the interstellar medium absorptions, the separation between both components is wide enough to allow a measurement of the carbon abundance.  The individual components of the \ion{C}{III} $\lambda$1175 sextuplet have an equivalent width of about 100 m\AA.  The detection of the \ion{C}{III} $\lambda$1175 sextuplet in both {\it FUSE} and STIS E140m spectra offers a way to verify that the additional broadening is not caused by unaccounted instrument motions. By using the sextuplet observed in both instruments, we measured almost identical C abundances, $v \sin i$, and microturbulent velocities $\xi$.

\subsubsection{Nitrogen}

Nitrogen lines are detected in the {\it FUSE}, STIS, and HIRES spectra. Three ionisation stages are observed:  \ion{N}{III}, \ion{N}{IV}, and \ion{N}{V}. All lines are in absorption except for three \ion{N}{III} emission lines that are observed in the HIRES spectrum, and the P~Cygni profiles in the \ion{N}{V} $\lambda$1240 doublet. We will discuss the P~Cygni profiles in more details in \S~\ref{mass_loss}.  By looking at Table~\ref{tab:atomic_data}, it is interesting to note that the photospheric absorptions in the \ion{N}{V} $\lambda$1240 doublet yield a significantly lower abundance than the ones corresponding to the other N lines. This lower abundance is likely due to the filling of the absorption lines by scattered light that comes from the stellar wind. The abundance of the \ion{N}{V} doublet is not included in the average abundance given in Table~\ref{tab:abundances}. 

To compute the \ion{N}{III} emission lines, we first used the \ion{N}{III} model atom that \cite{lanz_hubeny2003, lanz_hubeny2007}  built to compute their grids of non-LTE line-blanketed model atmospheres of B and O stars. The model atom contains 24 individual energy levels and considers 184 allowed transitions. Although our models predicted the appearance of the \ion{N}{III} emission lines $\lambda$4634 and 4640, they failed to produce the emission line $\lambda$4867. In fact, the models produced the $\lambda$4867 line in absorption. We therefore opted for a more elaborate \ion{N}{III} model atom, which we retrieved from the TLUSTY website\footnote{\url{http://nova.astro.umd.edu/}}. The model consists of 40 individual energy levels and 403 allowed transitions.  It includes the $\lambda$4867 $3p\ ^{4}{\rm{D}} - 3d\ ^{4}{\rm{F}}$ transition explicitly.  Using this model atom, we recomputed a grid of stellar models. The new models produced the \ion{N}{III} $\lambda$4634, 4640, and 4867 lines in emission. All of the N abundances given in Table~\ref{tab:atomic_data} are measured using the more sophisticated \ion{N}{III} model atom.   


The \ion{N}{IV} $\lambda$1036 and $\lambda$1078 lines also required additional effort.
The N abundances initially derived from these features were factors of 10 and 3 larger (respectively) than the mean.
We traced this discrepancy to our choice of oscillator strengths. We had used the atomic line lists that are provided on the TLUSTY website. These line lists were extracted from the compilation of  \citet{kurucz_bell95}\footnote{\url{http://kurucz.harvard.edu/linelists.html}} and updated with newer oscillator strengths and several new lines from the NIST Atomic Spectra Database\footnote{\url{http://www.nist.gov/pml/data/asd.cfm}}.  Although many \ion{N}{IV} lines have been updated, the $\lambda$1036 and $\lambda$1078 oscillator strengths definitely underestimated the strength of the observed lines. We replaced these oscillator strengths with the larger values found in Peter van Hoof's Atomic Line List website\footnote{\url{http://www.pa.uky.edu/~peter/atomic/}}, which were compiled from the Opacity Project \citep{tully_etal90}. The oscillator strengths agree with those published by \citet{allard_etal90, allard_etal91}, who compiled \ion{N}{IV} atomic data from several sources that include data from the Opacity Project. The $\lambda$1078 oscillator strength does not agree, however, with the value compiled by \citet{wiese96}, who report a lower value ($\log gf = 0.175$). Unfortunately, the \ion{N}{IV} $\lambda$1036 oscillator strength is not listed in the extensive compilation of \citet{wiese96}. Our observations favor the oscillator strengths that are reported in the compilations of Peter van Hoof and \citet{allard_etal90, allard_etal91}.  


\subsubsection{Oxygen}

Like nitrogen, oxygen is observed in three ionisation stages in the FUV, UV, and optical wavelength ranges. All lines are observed in absorption.  Although the abundances of the \ion{O}{IV} lines show a relatively large dispersion, the average \ion{O}{IV} abundance is consistent with the \ion{O}{III} and \ion{O}{V} abundances.

\subsubsection{Aluminum}

The Al abundance is determined by analysing the photospheric \ion{Al}{III} $\lambda$1854 and 1862 resonance doublet. Because the photospheric doublet is well separated from the interstellar medium component, its abundance is well determined. No other photospheric Al lines are detected. 

\subsubsection{Silicon}

A handful of \ion{Si}{III} and \ion{Si}{IV} transitions are observed. An absorption feature is observed at the wavelength corresponding to the photospheric \ion{Si}{III} $\lambda$1206 resonance line, but the feature seems to be blended. Because the nature of this additional opacity is unclear, the line is not included in the calculation of the abundance. The strong $\lambda$1206 interstellar medium feature is present at longer wavelengths. The \ion{Si}{III} $\lambda$1298 line is the only line of the sextuplet that is observed. 

%

\subsubsection{Sulfur}

Only one component of the \ion{S}{VI} resonance doublet is detected, because the $\lambda$933 line is blended with the \ion{He}{II} line ($n = 2 \rightarrow 13$). 

\subsubsection{Iron}

We determined the Fe abundance by fitting spectral regions that are dominated by only Fe lines. We selected three regions that include i) only \ion{Fe}{IV} lines, ii) only \ion{Fe}{V} lines, and iii) a mix of \ion{Fe}{IV} and \ion{Fe}{V} lines. Table~\ref{tab:atomic_data} identifies the lines that are included in these regions. It is worth noting that the Fe abundance that is determined when considering microturbulence is the same as the abundance derived without additional source of line broadening. Table~\ref{tab:abundances} shows that the Fe abundances are equal, although the abundances of the other elements are not. This illustrates the fact that the Fe lines are not saturated.  

\subsubsection{Nickel}

Only a few faint \ion{Ni}{IV} and \ion{Ni}{V} lines are detected. In fact, these lines are the ones that are expected to be observed, because they correspond to transitions with large oscillator strengths that start from low-lying energy levels.  No information on the projected stellar rotational velocity or microturbulent velocity could be obtained from these lines, because they are too faint. 

\section{Mass Loss}\label{mass_loss}

The FUV spectra of many hot post-AGB stars (particularly the central stars of planetary nebulae) exhibit wind features from a variety of species \citep[e.g.,][]{guerrero_demarco13}.  
The spectrum of vZ~1128 exhibits P~Cygni profiles in the \ion{N}{V} $\lambda 1240$ doublet.  Both components  
show broad blue-shifted absorption and faint red-shifted emission relative to the star's photospheric lines. 
The emission seems to be attenuated by interstellar \ion{N}{V} absorption. 

\begin{figure}
\includegraphics[width=87mm]{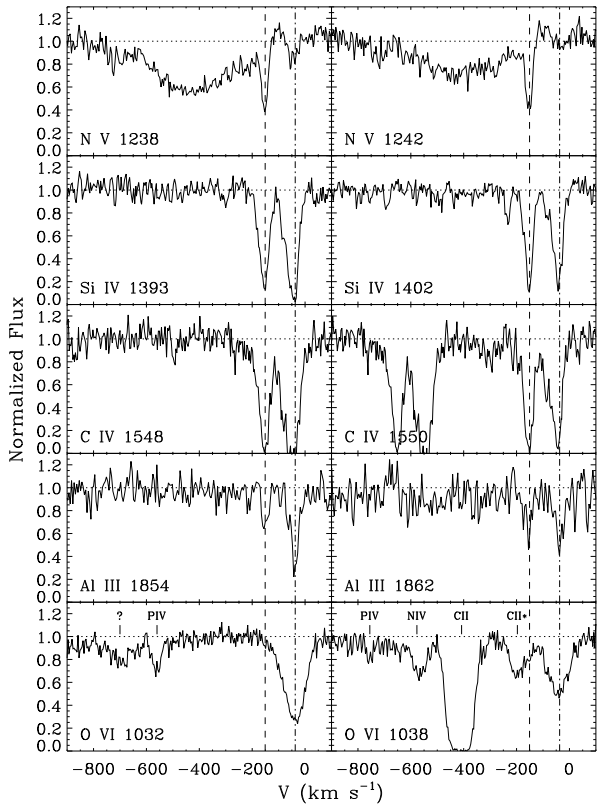}
\caption{Normalized profiles of the resonance doublets observed in the STIS and \fuse\/ spectra of vZ~1128 in the heliocentric frame. The dotted line corresponds to the normalized continuum. The vertical dashed line indicates the radial velocity of vZ~1128 ($V_{\rm{R}} = -150.9$ km$\,$s$^{-1}$). The vertical dot-dashed line indicates the central velocity of the interstellar \ion{O}{VI} lines, $V_{\rm{R}} =-37.2$ km$\,$s$^{-1}$, given in \citet{howk_sembach_savage03}. 
\label{fig:vr}}
\end{figure}

\fig{fig:vr} compares the \ion{N}{V} $\lambda$1240 lines to the resonance lines of \ion{Si}{IV} $\lambda$1397, \ion{C}{IV} $\lambda$1550, \ion{Al}{III} $\lambda$1857, and \ion{O}{VI} $\lambda$1034 observed in the STIS and \fuse\/ spectra.  For all five species, the photospheric and interstellar absorption features are aligned in velocity (with the exception of photospheric \ion{O}{VI} absorption, which is absent in this star).
The \ion{N}{V} P~Cygni absorption troughs span heliocentric velocities between about $-700$ km~s$^{-1}$ and $-150$ \kms\ and reach depths of $\sim 30$\% to 40\% below the continuum. The continuum for the other resonance lines is flat in this velocity range; no broad absorption is observed. The \ion{N}{V} $\lambda$1240 lines are the only spectral features in either the \fuse\/ or STIS bands that exhibit P~Cygni profiles.  

To constrain the properties of the wind, we fit a simple model to the \ion{N}{V} doublet using the line-fitting
routines of D.\ Massa, which combine the Sobolev with exact 
integration (SEI) algorithm described by \citet{lamers_etal87} with the
wind optical depth law introduced by \citet*{massa_etal95}.  
This approach uses the standard ``beta law'' parameterization of the velocity, which has the form
\begin{equation}
w(x) = w_0 + (1 - w_0)(1 - 1/x)^{\beta},
\end{equation}
where $w(x) = {v(x)}/{v_{\infty}}$, $x = r/R_*$, $R_*$ is the stellar
radius, and $w_0$ is the initial velocity at the base of the wind.
(In this formalism, all wind velocities are expressed as a fraction of the terminal velocity \vinf.)  The parameter $\beta$, which
defines the spatial velocity gradient in the wind, is
constrained by the shape of the emission component of the P-Cygni
profile.  
Other inputs to the code are the turbulent velocity $w_{\rm D}$, which simulates macroscopic velocity fields in the wind by smoothing the wind profile; the input photospheric spectrum; and  $\tau_{\rm r}(w)$, the radial component of the optical depth of the absorbing species as a function of the wind velocity.

\begin{figure}
\includegraphics[width=87mm]{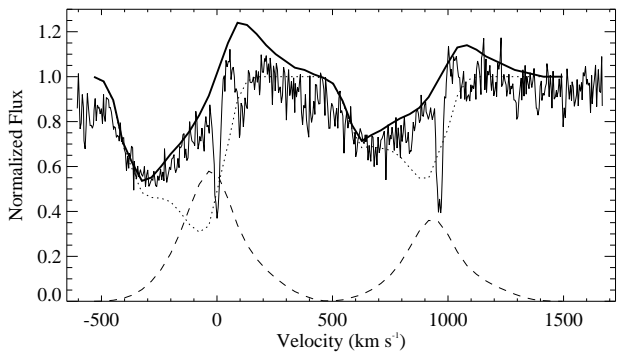}
\caption{N~{\sc v} $\lambda 1240$ doublet, which shows a P~Cygni profile. The data are overplotted by the wind model described in the text (thick solid line).  Also shown are the transmitted (dotted line) and scattered (dashed line) components of the model profile. The spectrum is plotted in the stellar rest frame and centered on the blue component of the doublet.  Photospheric and interstellar nitrogen absorption features are not included in the model.
\label{fig:fit_pcygni}}
\end{figure}

Once $\beta$ and $v_{\infty}$ are determined, the resonance lines are fit by adjusting $w_D$ and the elements of $\tau_{\rm r}(w)$.  The photospheric spectrum is modeled with a flat continuum.  The \ion{N}{V} profile is well fit by the parameters $v_{\infty} = 380$ \kms, $w_0 = 0.01$, $\beta = 0.5$, and $w_{\rm D} = 0.2$.  The optical depth is roughly constant throughout the wind, with $\tau_{\rm r} \sim 0.8$.  Our best-fit model is presented in \fig{fig:fit_pcygni}.  A complete explanation of this technique, its applicability, and its limitations may be found in \citet{massa_etal03}.

From the stellar abundances and our wind model, we can estimate $\dot{M}q(w)$, the product of the stellar mass-loss rate and the fraction of nitrogen in the form of N$^{+4}$ as a function of velocity in the wind.  Averaging over the velocity range $0.2 < w < 0.75$, where our fit to $\tau_{\rm r}$ is most reliable, we find that $\langle \dot{M}q \rangle = 10^{-11}$ \msun\ yr$^{-1}$.  Our stellar model predicts that $q = 0.1$ at the outer limit of the photosphere.  If $q$ is constant throughout the wind, then $\dot{M} = 10^{-10}$ \msun\ yr$^{-1}$.
\citet{pauldrach88} compute terminal velocities and mass-loss rates
for the winds of Pop.~I central stars of PNe.  For a star with $M =
0.546 \; M_{\sun}$ (\S \ref{evolution}) and \teff\ = 36,000 K, their
models predict $v_{\infty} \sim 1700$ \kms\ and  $\dot{M} \sim 1.5 \times 10^{-9}$ $M_{\sun}$ yr$^{-1}$.
It is likely that models assuming Pop.~II abundances would predict lower values for both parameters, since such stars have fewer opacity sources to drive the wind.  It is also likely that $q(w)$ is not constant, but falls with increasing velocity (and thus distance from the star), so that our \ion{N}{V} profile does not probe the highest velocities achieved by the wind.

In the photosphere of vZ~1128, the abundances of nitrogen and oxygen exceed those of the other metals by more than an order of magnitude.  The star is too cool to exhibit significant \ion{O}{VI} absorption.  Should we expect wind features from other species?  If two ions, $1$ and $2$, exist throughout the volume of the wind, then their relative optical depths depend on the run of ionization fractions $q(r)$, the abundance of each element $A$, the relative oscillator strengths of the transition $f$, and their wavelengths:
\begin{equation}
{\tau_1 \over \tau_2} = {q_1(r)\,A_1\,{f_1\,\lambda_1} \over {q_2(r)\,A_2}\, f_2\,\lambda_2}\enspace.
\label{eq_ratio}
\end{equation}
If their ionization fractions are constant throughout the wind at their photospheric values, then the relative strengths of the wind features due to \ion{N}{V}, \ion{Si}{IV}, and \ion{C}{IV} are presented in Table~\ref{compare}.  Based on this simple analysis, we would not expect detectable wind features from either \ion{Si}{IV} or \ion{C}{IV}.

\begin{figure}
\includegraphics[width=87mm]{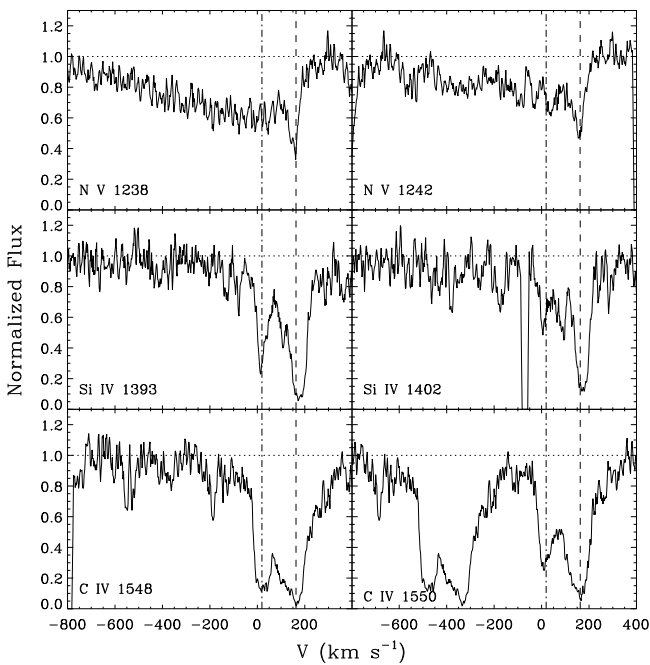}
\caption{Normalized profiles of resonance doublets in the COS spectrum of ELS 31 in the SMC cluster NGC 346, plotted in a heliocentric reference frame.  The dotted line corresponds to the normalized continuum. The vertical dashed lines indicate the radial velocity of the star (162.9 km s$^{-1}$). The vertical dot-dashed lines indicate the central velocity of the interstellar features due to our galaxy (19.2 km s$^{-1}$).  Like vZ~1128, the star exhibits P~Cygni profiles only in the \ion{N}{V} $\lambda$1240 doublet.
\label{fig:NGC346}}
\end{figure}

Effects other than abundance may be at work.  \citet{bouret_etal13} observed 22 main-sequence O-type stars in the SMC with the Cosmic Origins Spectrograph (COS) aboard \hst.  Several of these stars have effective temperatures and surface gravities similar to those of vZ~1128, but they are considerably more metal rich, with abundances roughly a factor of five below solar.  In particular, their atmospheres are not enhanced in nitrogen.  Nevertheless, they exhibit the pattern seen in vZ~1128: broad P~Cygni profiles in both components of the \ion{N}{V} $\lambda$1240 doublet, but no other FUV wind features.  As an example, \fig{fig:NGC346} presents the \ion{N}{V} $\lambda$1240, \ion{Si}{IV} $\lambda$1397, and \ion{C}{IV} $\lambda$1550 features of the star ELS 31 ($T_{\rm eff} = 37,200$K, $\log g = 4.00$) in the star cluster NGC~346 \citep{evans_etal06}.  Because the star has a C/N ratio of 4.0, Eq.~\ref{eq_ratio} predicts that the optical depth of wind features in the \ion{C}{IV} doublet would be 2.5 times those in \ion{N}{V}, yet no blue-shifted \ion{C}{IV} absorption is seen.

\begin{table}
\caption{Optical Depths of Possible Wind Features}
\begin{tabular}{cccc}
\hline\hline\noalign{\smallskip}
Parameter & N~{\scshape v} & Si~{\scshape iv} & C~{\scshape iv}  \\
\hline\noalign{\smallskip}
$q(r=R_\star)$           &      0.10          &    $<0.01$         &          0.04       \\
$\log A/H$                     &    $-$4.44       & $-$6.03         &  $-$5.59         \\
$f$                               &    0.156         &  0.513              &   0.190       \\
$\lambda$ (\AA)                 &  1238.821     &   1393.755     &    1548.202    \\
$\tau_r / \tau_r$(\ion{N}{V}) &    1.00           & $< 0.01$       & 0.05                \\
\hline\noalign{\smallskip}
\end{tabular}
\label{compare}
\end{table}

\begin{figure}
\includegraphics[width=87mm]{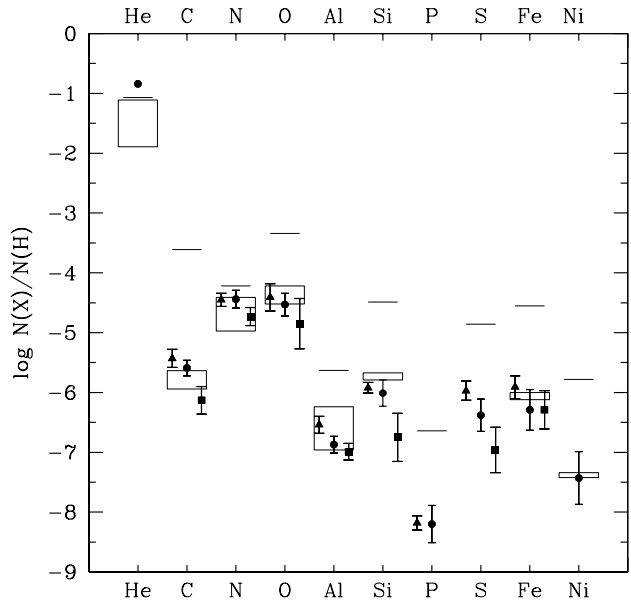}
\caption{Comparison of the abundances derived for vZ~1128 (filled symbols) with those of the solar photosphere (short horizontal lines) and of HB and RGB stars in M3 (rectangles).  Stellar abundances derived without an additional source of line broadening are plotted as circles, those allowing for stellar rotation are plotted as triangles, and those allowing for microturbulence are plotted as squares.\label{fig:abun_abs}}
\end{figure}

\begin{figure}
\includegraphics[width=85mm]{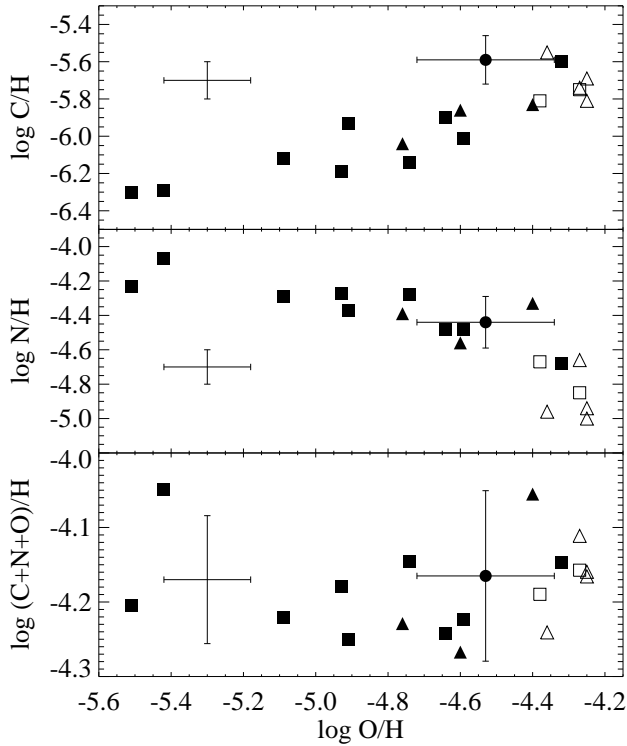}
\caption{Abundance ratios of red giants in the globular clusters M3 (triangles) and M13 (squares).  CN-rich stars are plotted as solid symbols, CN-poor stars as open symbols.  The abundances of C and O are correlated, N and O are anticorrelated, and the total abundance of (C+N+O) is essentially constant, consistent with the products of CNO-cycle processing.  The abundance ratios of vZ~1128 in M3 (circles) are consistent with the patterns seen in the RGB stars.
\label{fig:smith96}}
\end{figure}

\section{Discussion}\label{sec_discussion}

In \fig{fig:abun_abs}, the measured abundances of vZ~1128 (solid symbols) are compared with those of the sun (short horizontal lines; \citealt{asplund2009}) and M3 (rectangles).  The He abundance is measured on the horizontal branch \citep[HB;][]{behr_etal03}.  Metal abundances are from red-giant-branch (RGB) stars: the C and N abundances are from \citet{smith_etal96}, and the O, Al, Si, Fe, and Ni values are from \citet{sneden_etal04}.  The vertical extent of each rectangle represents the star-to-star scatter in the measured abundance ($\pm 1 \sigma$ about the mean).  Beginning with the most massive elements, we see that the abundances of Si, Fe, and Ni are nearly constant along the RGB.  The scatter is much larger for CNO and Al, reflecting well-known abundance variations in globular-cluster giants \citep{kraft94}.  For all of these elements, the measured abundances of vZ~1128 are consistent with those of the RGB stars. 

\subsection{Helium Abundance}\label{sec_helium}

The He abundance of vZ~1128 is nearly twice the solar value \citep[$\log N({\rm{He}})/N({\rm{H}}) = -1.07$;][]{asplund2009}.  While RGB stars are too cool to show helium absorption, HB stars with 9000 K $<$ \teff\ $<$ 11,200 K are hot enough to excite optical \ion{He}{I} lines but not so hot that gravitational settling reduces the surface helium abundance.  \citet{behr_etal03} derived helium abundances for five HB stars in M3. Though their uncertainties are large, they average $\log N({\rm{He}})/N({\rm{H}}) = -1.50 \pm 0.39$, roughly a third of the solar value.  If this result is borne out, then vZ~1128 would be enhanced in He relative to the cluster mean.  

\subsection{CNO Abundances} \label{sec_cno}

\citet{smith_etal96} investigated the variations in CNO abundance of stars on the RGB of M3 and M13, which have similar metallicities.  We plot their data in \fig{fig:smith96}, with the following changes: we express abundances in terms of $\log N({\rm{X}})/N({\rm{H}})$ rather than [X/Fe], and we use oxygen abundances from \citet{sneden_etal04} where available.  
The results of \citet{smith_etal96} are reproduced: the abundances of C and O are correlated, the N abundance is anticorrelated with both C and O, and the total abundance C+N+O is nearly constant.  These patterns can be explained as the result of CNO-cycle hydrogen burning, which converts carbon (rapidly) and oxygen (slowly) into nitrogen, but leaves the total C+N+O abundance unchanged.  Comparing abundances derived from non-LTE models of a sdO-type star observed in the FUV with those derived from LTE models of K-type giants observed in the optical may not be appropriate; nevertheless, we have added vZ~1128 to \fig{fig:smith96}.  Though its carbon abundance is a bit high, the star's CNO abundances follow the trends seen in the cluster's RGB stars remarkably well.  (Note that we plot the vZ~1128 abundances derived without an additional source of line broadening; these values are plotted as circles in \fig{fig:abun_abs}.)

\subsection{Iron Abundance}

Spectroscopic analyses of hot post-AGB field stars at high Galactic latitudes (\citealt{mccausland_etal92}, \citealt{napiwotzki_heber_koeppen92}) reveal that their Fe abundances are quite low, on the order of [Fe/H] = $-2.0$.   \citet{moehler_etal98} found that, among hot post-AGB stars in globular clusters, Barnard~29 in M13 and ROA~5701 in $\omega$
Cen exhibit iron depletions of $\sim 0.5$ dex relative to the cluster RGB.  Changes in iron abundance due to nucleosynthesis are not expected in low-mass stars, so these results are puzzling.   Using new optical data, \citet{thompson_etal07} have recently derived iron abundances for both Barnard~29 and ROA~5701 that are consistent with the cluster mean.  We obtain a similar result for vZ~1128 (\fig{fig:abun_abs}). 

\citet{dixon_davidsen_ferguson94} quote a metallicity for vZ~1128 of [M/H] = $-3.5 \pm 1.5$.  In their analysis, the authors compared the shape of the star's far-UV spectrum (900--1850 \AA) with those of synthetic spectra computed by \citet{kurucz92}, which assume solar abundance ratios.  The authors reported that models with abundances [$-2.5$], [$-3.0$], and [$-3.5$] all fell within the $2 \sigma$ error ellipse of the best-fit model.  While the grid of models with abundance [$-2.0$] did not extend to temperatures above 32,000 K, they estimated that models with this abundance would also have fallen within the error ellipse.  No models with abundance of [$-1.5$] or higher fell within the error ellipse.  In effect, \citet{dixon_davidsen_ferguson94} were able to set only an upper limit of [M/H] = $-1.5$ on the star's metallicity, a value consistent with our iron abundance, which is derived from fits to individual spectral features.

\subsection{Stellar Evolution}\label{evolution}

A star's atmospheric abundances are modified by its post-main sequence evolution.  According to stellar evolutionary theory \citep{iben_renzini83}, a star's arrival on the red giant branch is accompanied by a deepening of its convective envelope, which brings to the surface the ashes of hydrogen burning via the CNO cycle.  As a result of this process, known as first dredge-up, we expect to see a doubling of the surface $^{14}$N abundance, a reduction in the $^{12}$C abundance of about 30\%, and practically no change in the abundance of $^{16}$O.  

No further mixing on the RGB is predicted by theory, but observations of low-mass red giants in both clusters and the field \citep[e.g.,][]{gratton_etal00} suggest that some form of mixing continues to bring CNO-processed material to the surface.  The carbon abundance continues to fall, and the nitrogen abundance to rise (corresponding to a migration from right to left in \fig{fig:smith96}) as a star ascends the RGB \citep{smith02}.  

No mixing is expected to occur on the horizontal branch.  A second period of dredge-up can occur at the base of the AGB, but only in stars more massive than those seen in globular clusters today.  Third dredge-up occurs near the tip of the AGB, as stars experience thermal pulses in their helium-burning shells.  Third dredge-up brings $^{12}$C and s-process isotopes produced by the burning of helium into the convective envelope.  As a result, the atmospheric C/O ratio quickly exceeds unity \citep{van_winckel03}.  We expect vZ~1128 to exhibit the effects of first dredge-up, any extra mixing that took place on the RGB, and third dredge-up, if it occurred.  

The stellar temperature and luminosity derived by \citet{dixon_davidsen_ferguson94} place vZ~1128 on the 0.546 $M_{\sun}$ post-AGB evolutionary track of \cite{schoenberner83}.  This track traces the evolution of a star that leaves the AGB before the onset of thermal pulsing.  Such objects are known as post-early AGB (post-EAGB) stars.  This scenario is consistent with the abundance pattern seen in the atmosphere of vZ~1128: its carbon abundance is not enhanced, nor are any s-process elements detected.  We conclude that the star has not undergone third dredge-up.  Indeed, it appears that no significant changes in its atmospheric composition have occurred since the star left the RGB.

We should not be surprised that vZ~1128 is a post-EAGB star.  Post-AGB stars evolve quickly, remaining luminous for only $10^3$--$10^4$ years, while post-EAGB stars remain luminous for $10^4$--$10^5$ years \citep{schoenberner81, schoenberner83}.  Of the roughly one dozen UV-bright stars in globular clusters whose spectra have been analyzed to date (see \citealt{moehler10} for a review), only two show the enhanced carbon abundance expected of a star that evolved to the tip of the AGB.  The first, K648 in M15 \citep{rauch_heber_warner02}, hosts a planetary nebula.  The second, ZNG~1 in M5, lacks a nebula, and its high helium abundance and high rotational velocity suggest an unusual evolutionary history \citep{dixon_etal04}.  The dearth of carbon-rich post-AGB stars in galactic globular clusters is consistent with the short lifetimes of these rare objects.


\section{Conclusions}\label{sec_summary}

We have analyzed \fuse, STIS, and Keck spectra of the UV-bright star vZ~1128 in M3. 
We determine the star's atmospheric parameters by fitting the hydrogen and helium lines in its Keck spectrum with non-LTE H-He models. The star's \fuse\/ and STIS spectra show photospheric absorption from C, N, O, Al, Si, P, S, Fe, and Ni, but no absorption from elements beyond the iron peak.  We determine the abundance of each element by comparing its observed lines to a grid of synthetic spectra.  Additional broadening is required to reproduce the line cores, though we cannot distinguish between stellar rotation or microturbulence as the source.  Modeling the star's nitrogen lines proved particularly challenging:   Reproducing three \ion{N}{III} emission features in the star's HIRES spectrum required the use of a more sophisticated model atom.  Obtaining reasonable nitrogen abundances from a pair of \ion{N}{IV} absorption features in the star's \fuse\/ spectrum required the use of larger oscillator strengths.  Finally, both components of the \ion{N}{V} $\lambda 1240$ doublet in the star's STIS spectrum exhibit P~Cygni profiles, indicating a weak stellar wind, though no other wind features are seen.  The star's photospheric abundances appear to have changed little since it left the RGB.  Its C, N, O, Al, Si, Fe, and Ni abundances are consistent with those of cluster RGB stars; in particular, the relative abundances of C, N, and O follow the trends seen on the cluster RGB.  Its low C abundance suggests that the star left the AGB before the onset of third dredge-up, making vZ~1128 a post-early AGB star.

\section*{Acknowledgments}

The authors wish to thank L. M. Torres and J. M. Rivera for their assistance with spectral-line fitting, J.~C.~Howk for help with the STIS data, and I.~Hubeny and T.~Lanz for providing their stellar atmosphere and spectral synthesis codes.  This research has made use of NASA's Astrophysics Data System Bibliographic Services and the SIMBAD database, operated at CDS, Strasbourg, France.  It has also made use of the Keck Observatory Archive (KOA), which is operated by the W.~M. Keck Observatory and the NASA Exoplanet Science Institute (NExScI), under contract with NASA. The STIS data presented in this paper were obtained from the Mikulski Archive for Space Telescopes (MAST).  STScI is operated by the Association of Universities for Research in Astronomy, Inc., under NASA contract NAS5-26555.   IDL is a registered trademark of Exelis Visual Information Solutions, Inc., for its Interactive Data Language software.  This work was supported by NASA grant NAS5-32985 to the Johns Hopkins University. P.C. is supported by the Canadian Space Agency under a contract with NRC Herzberg Astronomy and Astrophysics.  L. M. Torres and J. M. Rivera were supported by NASA contract NNG04D58G. 



\bibliographystyle{mn2e}
\bibliography{vz1128}









%
\end{document}